\documentclass[twocolumn]{aastex63}
\usepackage{amssymb,amsmath}
\usepackage{graphicx}
\usepackage{xcolor}
\usepackage{multirow}
\usepackage[T1]{fontenc}
\usepackage{ae,aecompl}
\usepackage{newtxtext,newtxmath}
\usepackage{longtable,listings}

\def\oii{[O~{\sc ii}]$\lambda3727$\AA\ }

\submitjournal{ApJ}

\shorttitle{BH mass in SDSS J0159}
\shortauthors{Zhang XueGuang}

\begin{document}

\title{Central BH mass of tidal disruption event candidate SDSS J0159 through long-term optical variabilities}

\correspondingauthor{XueGuang Zhang}%
\email{xgzhang@gxu.edu.cn}
\author{XueGuang Zhang$^{*}$}
\affiliation{Guangxi Key Laboratory for Relativistic Astrophysics, School of Physical Science and Technology,
GuangXi University, No. 100, Daxue Road, Nanning, 530004, P. R. China}

\begin{abstract} 
	In this manuscript, central BH mass is determined in the tidal disruption event (TDE) candidate SDSS 
J0159, through the nine years long variabilities, in order to check whether the virial BH mass is consistent 
with the mass estimated by another independent methods. First, host galaxy spectroscopic features are described 
by 350 simple stellar templates, to confirm the total stellar mass about $7\times10^{10}{\rm M_\odot}$ in SDSS 
J0159, indicating the virial BH mass about two magnitudes larger than the BH mass estimated by the total stellar 
mass. Second, based on an efficient method and fitting procedure, through theoretical TDE model applied to 
describe the SDSS $ugriz$-band light curves of SDSS J0159, central BH mass can be determined as 
$M_{BH}\sim4.5_{-1.1}^{+1.3}\times10^6{\rm M_\odot}$, well consistent with the M-sigma relation expected BH 
mass and the total stellar mass expected BH mass. Third, the theoretical TDE model with parameter of central 
BH mass limited to be higher than $10^8{\rm M_\odot}$ can not lead to reasonable descriptions to the light curves 
of SDSS J0159, indicating central BH mass higher than $10^8{\rm M_\odot}$ is not preferred in SDSS J0159. 
Therefore, the TDE model determined central BH mass of SDSS J0159 are about two magnitudes lower than the 
virial BH mass, to support central BLRs including accreting debris contributions from central TDE, and provide 
interesting clues to reconfirm that outliers in the space of virial BH mass versus stellar velocity dispersion 
should be better candidates of TDE.
\end{abstract}

\keywords{
active galactic nuclei -- emission line galaxies -- supermassive black holes -- tidal disruption -- 
transient sources
}

\section{Introduction}
 
	SDSS J0159 (=SDSS J015957.64+003310.5) at redshift $z=0.312$ (corresponding luminosity distance about 
1625Mpc) is an interesting object in the literature. \citet{lc15} have reported the SDSS J0159 as the first 
changing-look quasar transitioned from a Type 1 quasar (both apparent broad H$\alpha$ and H$\beta$ in optical 
spectrum) to a Type 1.9 AGN (Active Galactic Nuclei) (only weak broad H$\alpha$ in optical spectrum) between 
2000 and 2010, and reported the dimming of the AGN continuum as intrinsic physical reason to explain the 
changing-look properties. Meanwhile, \citet{md15} have reported the SDSS J0159 as a candidate of Tidal Disruption 
Event (TDE) due to the very rapid rise and the decay trend of the long-term optical variabilities well described 
by $t^{-5/3}$ expected by theoretical TDE model, indicating TDEs can be well treated as one explanation to 
changing-look AGN as discussed in \citet{yw18, zh21b}.

	A star can be tidally disrupted by gravitational tidal force of a central massive black hole (BH), 
when it passing close to the central BH with a distance larger than event horizon of the BH but smaller 
than the expected tidal disruption radius $R_{{\rm T}}~=~R_\star~\times~(\frac{M_{{\rm BH}}}{M_{\star}})^{1/3}$, 
where $R_\star$, $M_{\star}$ and $M_{{\rm BH}}$ represent radius and mass of the tidally disrupted star 
and central BH mass, respectively. And then, fallback materials from tidally disrupted star can be 
accreted by the central massive BH, leading to a flare up follows by a decrease. This is the basic picture 
of a TDE.

	The well-known pioneer work on TDE can be found in \citet{re88}. Since then, as an excellent 
beacon for indicating massive black holes, both theoretical simulations and observational results on 
TDEs have been widely studied and reported in the literature. More detailed and improved simulations on 
TDEs can be found in \citet*{ek89, MT99, be04, lk09, mg12, gr15, lf15, bp17, dc18, cer19, cn19, gc19}, 
etc.. More recent reviews in detail on TDEs can be found in \citet{st18}. And based on theoretical TDE 
model, public codes of {\it TDEFIT} provided by \citet{gm14} (\url{https://github.com/guillochon/tdefit}) 
and {\it MOSFIT} provided by \citet{gn18} (\url{https://github.com/guillochon/mosfit}) have been well 
applied to describe both structure evolutions in details for the falling back stellar debris and evolutions 
of expected time-dependent long-term variability through hydrodynamical simulations. Therefore, TDE 
models with sufficient model parameters can be applied to describe observed long-term variability from 
TDEs.

	Meanwhile, there are so-far around one hundred observational results on TDE candidates reported 
in the literature based on TDE expected variability properties in different multi-wavelength bands, 
such as the reported TDE candidates in \citet{kh04, ce12, gs12, ho14, ho16, lg17, ts17, mp18, wt18, 
ha20, ll20, nh20, gv22}, etc.. More recently, \citet{vg21} have reported seventeen TDE candidates from 
the First Half of ZTF (Zwicky Transient Facility) Survey observations along with Swift UV and X-ray 
follow-up observations. \citet{sg21} have reported thirteen TDE candidates from the SRG all-sky survey 
observations and then confirmed by follow-up optical observations. More recent review on observational 
properties of reported TDEs can be found in \citet{gs21}.

	Among the reported TDE candidates, SDSS J0159 is an interesting object, because its central 
virial BH mass reported as $\sim10^8{\rm M_\odot}$ in \citet{lc15, md15} through the virialization 
assumptions \citep{pe04, gh05, vp06, kb07, rh11, sh11, mt22} applied to BLRs (broad emission line regions) 
clouds, the largest BH mass among the reported BH masses of TDE candidates in \citet{wv17, mg19, zl21, 
wp22}. However, variations of accretion flows have apparent effects on dynamical structures of BLRs, if 
the BLRs clouds are tightly related to TDEs, such as the more detailed abnormal variabilities of broad 
emission lines in the TDE candidate ASASSN-14li in \citet{ho16}: strong emissions leading to wider line 
widths of broad H$\alpha$ which are against the expected results by the Virialization assumptions to BLRs 
clouds. More recently, \citet{zh19} have measured the stellar velocity dispersion about $81\pm27{\rm km/s}$ 
of SDSS J0159 through the absorption features around 4000\AA, reported that the M-sigma relation 
\citep{fm00, ge00, kh13, bt15} determined central BH mass are about two magnitudes smaller than the virial 
BH mass in SDSS J0159, and provided an interesting clue to detect TDE candidates by outliers in the space 
of virial BH masses versus stellar velocity dispersions.

	Moreover, \citet{cr19} have shown the total stellar mass of SDSS J0159 is about 
$4.7\times10^{10}{\rm M_\odot}$, indicating the central BH mass about $6\times10^{6}{\rm M_\odot}$ (the 
accepted value in \citet{wp22}) in SDSS J0159 through the correlation between central BH mass and total 
stellar mass as discussed in \citet{hr04, san11, kh13, rv15}, roughly consistent with the BH mass 
estimated by the M-sigma relation reported in our previous paper \citet{zh19}. Therefore, besides 
the virial BH mass through virialization assumption to BLRs clouds and the BH mass through the M-sigma 
relation and through the total stellar mass, BH mass of SDSS J0159 determined through one another 
independent method will be very interesting and meaningful.

	More recently, \citet{gm14, mg19, rk20, zl21} have shown that the long-term TDE model expected 
time-dependent variability properties can be well applied to estimate central BH masses of TDE 
candidates. However, SDSS J0159 is not discussed in \citet{mg19, rk20, zl21}, probably due to lack of 
information of peak intensities of light curves in SDSS J0159 and/or other unknown reasons. Therefore, 
in the manuscript, the central BH mass of SDSS J0159 is to be estimated by theoretical TDE model applied 
to describe the long-term optical variabilities. It is very interesting to check whether would the 
reported virial BH mass or M-sigma relation determined BH mass be coincident with the TDE model determined 
BH mass. The manuscript is organized as follows. In section 2, we present spectroscopic features of 
SDSS J0159, in order to re-confirm the low stellar velocity dispersion and the low total stellar mass. 
Section 3 shows our methods with a TDE model applied to describe observed multi-band light curves. 
Then, the main results and necessary discussions are shown in Section 4. Section 5 shows simulating 
results to check whether the long-term variabilities are tightly related to a central TDE in SDSS J0159. 
Section 6 gives our main summary and final conclusions. And in the manuscript, the cosmological parameters 
of $H_{0}~=~70{\rm km\cdot s}^{-1}{\rm Mpc}^{-1}$, $\Omega_{\Lambda}~=~0.7$ and $\Omega_{m}~=~0.3$ 
have been adopted.

\begin{figure}
\centering\includegraphics[width = 8cm,height=6cm]{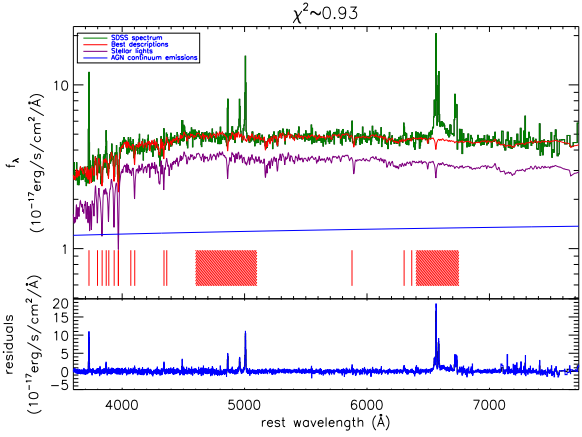}
\caption{The SDSS spectrum (solid dark green line) of SDSS J0159 and the best descriptions (solid red line) 
to the spectrum with the emission lines being masked out by the SSP method with applications of 350 stellar 
templates. In top region, as shown legend in top-left corner, solid purple line shows the SSP 
method determined stellar lights, and solid blue line shows the determined power law AGN continuum emissions, 
vertical red lines from left to right mark the following emission features masked out, including \oii, 
H$\theta$, H$\eta$, [Ne~{\sc iii}]$\lambda3869$\AA, He~{\sc i}$\lambda3891$\AA, Calcium K line, 
[Ne~{\sc iii}]$\lambda3968$\AA, Calcium H line, [S~{\sc ii}]$\lambda4070$\AA, H$\delta$, H$\gamma$, 
[O~{\sc iii}]$\lambda4364$\AA, He~{\sc i}$\lambda5877$\AA\ and [O~{\sc i}]$\lambda6300,6363$\AA, respectively, 
and the area filled by red lines around 5000\AA\ shows the region masked out including the emission features 
of probable He~{\sc ii}, broad and narrow H$\beta$ and [O~{\sc iii}] doublet, and the area filled by red 
lines around 6550\AA\ shows the region masked out including the emission features of broad and narrow 
H$\alpha$, [N~{\sc ii}] and [S~{\sc ii}] doublets. Bottom region shows the residuals calculated by SDSS 
spectrum minus sum of the stellar lights and the power law continuum emissions. 
}
\label{spec}
\end{figure}

\section{Spectroscopic results of SDSS J0159}

	In \cite{zh19}, the stellar velocity dispersion about $81\pm27{\rm km/s}$ of SDSS J0159 has been 
measured through absorption features around 4000\AA~ in the SDSS spectrum with plate-mjd-fiberid=3609-55201-0524 
which includes apparent host galaxy contributions, based on the 39 simple stellar population templates from 
the \citet{bc03}. In the section, the SSP (simple Stellar Population) method discussed in \citet{ref33, kh03, 
ca04, cf05, cp13, zh14, lc16, ca17, wc19} is re-applied to mainly determine the total stellar mass in SDSS 
J0159 but with more plenty of stellar templates, in order to check whether are there low total stellar mass 
in SDSS J0159, which will provide further clues on central BH mass. In the section, the spectrum with 
plate-mjd-fiberid=3609-55201-0524 is mainly considered and collected from the eBOSS (Extended Baryon Oscillation 
Spectroscopic Survey) \citep{ap20, ak22}, due to the following main reason. The eBOSS spectrum with apparent 
stellar absorption features of SDSS J0159 in the Type 1.9 AGN state is observed at the late times of the TDE 
expected flare, with few effects of TDE on the spectroscopic features.

	Due to the weak but apparent broad H$\alpha$ in SDSS J0159 observed in mjd=55201, contributions from 
central AGN continuum emissions should be considered. Here, 350 SSP templates $S_{lib}$ are collected from 
the MILES (Medium resolution INT Library of Empirical Spectra) stellar library \citep{ref34, ks21} with 50 
stellar ages from 0.06Gyrs to 17.78Gyrs and with 7 metallicities from -2.32 to 0.22, and the simple SSP 
method is re-applied to determine the host galaxy stellar lights and the central AGN continuum emissions, 
through the following model function
\begin{equation}
S_{g}~=~A~\otimes~S_{lib, \sigma, V_s}~+~\alpha\times\lambda^{\beta}
\end{equation}
with $S_g$ as the SDSS spectrum of SDSS J0159, $S_{lib, \sigma, V_s}$ as the broadened and shifted stellar 
templates with broadening velocity $\sigma$ and shifting velocity $V_s$, $\alpha\times\lambda^{\beta}$ as the 
AGN continuum emissions. Then, through the Levenberg-Marquardt least-squares minimization technique (the known 
MPFIT package), with emission lines being masked out, the weights $A$ and the power law continuum emissions 
can be well determined. Fig.~\ref{spec} shows the best descriptions and corresponding residuals to the SDSS 
spectrum with rest wavelength range from 3650\AA~ to 7700\AA~ by the model function above, with $\chi^2\sim0.93$ 
(summed squared residuals divided by degree of freedom) and with residuals calculated by SDSS spectrum minus 
the best descriptions.

	Based on the best descriptions, total stellar mass is determined to be about 
\begin{equation}
M_{tot}~\sim~\sum~A~\times~4\pi~Dis^2./Unit/L_{\odot}\sim7.11\times10^{10}{\rm M_\odot}
\end{equation}
with $L_{\odot}=3.826\times10^{33}{\rm erg/s}$ as the solar luminosity, and $Dis=1623Mpc$ as the distance 
between SDSS J0159 and the earth, and $Unit=10^{-17}{\rm erg/s/cm^2/\textsc{\AA}}$ as the emission intensity 
unit of the SDSS spectrum. The calculated total stellar mass in SDSS J0159 is roughly consistent with the 
reported $4.7_{-3.9}^{+8.0}\times10^{10}{\rm M_\odot}$ in \citet{cr19} and the reported 
$2.3_{-0.1}^{+0.7}\times10^{10}{\rm M_\odot}$ in \citet{gf18}, indicating lower central BH mass in SDSS J0159.

	Therefore, the low total stellar mass in SDSS J0159 can be well confirmed, and lead to expected 
lower central BH mass than the Virial BH mass in SDSS J0159. Then, besides the lower total stellar mass in SDSS 
J0159, it is interesting to check whether the TDE model can still lead to different central BH mass in SDSS 
J0159.

\section{Main Method and fitting procedure to describe the light curves}

	Similar as what we have recently done in \citet{zh22} to describe X-ray variabilities in the TDE 
candidate {\it Swift} J2058.4+0516 with relativistic jet, the following four steps are applied to describe 
the long-term optical $ugriz$-band variabilities of SDSS J0159. The similar procedures have also been applied 
in \citet{zh22b} to discuss TDE expected long-term variabilities in the high redshift quasar SDSS J014124+010306 
and in \citet{zh22c} to discuss TDE expected long-term variabilities of broad H$\alpha$ line luminosity in 
the known changing-look AGN NGC 1097.

	First, standard templates of viscous-delayed accretion rates in TDEs are created. Based on the 
discussed $dM/dE$ provided in \citep{gm14, gn18, mg19} (the fundamental elements in the public 
code of TDEFIT and MOSFIT), templates of fallback material rate $\dot{M}_{fbt}=dM/dE~\times~dE/dt$ can be 
calculated with $dE/dt~\sim~\frac{(2~\pi~G~M_{\rm BH})^{2/3}}{3~t^{5/3}}$, for the standard cases with 
central BH mass $M_{\rm BH}=10^6{\rm M_\odot}$ and disrupted main-sequence star of $M_{*}=1{\rm M_\odot}$ 
and with a grid of the listed impact parameters $\beta_{t}$ in \citet{gr13}. Considering the viscous delay 
effects as discussed in \citet{gr13, mg19} by the viscous timescale $T_{vis}$, templates of viscous-delayed 
accretion rates $\dot{M}_{at}$ can be determined by
\begin{equation}
\dot{M}_{at}~=~\frac{exp(-t/T_{vis})}{T_{vis}}\int_{0}^{t}exp(t'/T_{vis})\dot{M}_{fbt}dt'
\end{equation}.
Here, a grid of 31 $\log(T_{vis,~t}/{\rm years})$ range from -3 to 0 are applied to create templates 
$\dot{M}_{at}$ for each impact parameter. The final templates of $\dot{M}_{at}$ include 736 (640) 
time-dependent viscous-delayed accretion rates for 31 different $T_{vis}$ of each 23 (20) impact parameters 
for the main-sequence star with polytropic index $\gamma$ of 4/3 (5/3).

	Second, for common TDE cases with model parameters of $\beta$ and $T_{vis}$ different from the list 
values in $\beta_{t}$ and in $T_{vis,~t}$, the actual viscous-delayed accretion rates $\dot{M}_{a}$ are 
created by the following two linear interpolations. Assuming that $\beta_1$, $\beta_2$ in the $\beta_{t}$ 
are the two values nearer to the input $\beta$, and that $T_{vis1}$, $T_{vis2}$ in the $T_{vis,~t}$ are the 
two values nearer to the input $T_{vis}$, the first linear interpolation is applied to find the viscous-delayed 
accretion rates with input $T_{vis}$ but with $\beta=\beta_1$ and $\beta=\beta_2$ by
\begin{equation}
\begin{split}
&\dot{M}_{a}(T_{vis},~\beta_{1})=\dot{M}_{at}(T_{vis1},~\beta_1) + \\
&\ \ \ \frac{T_{vis}-T_{vis1}}{T_{vis2}-T_{vis1}}(\dot{M}_{at}(T_{vis2},~\beta_1)
	- \dot{M}_{at}(T_{vis1}, \beta_1))\\
&\dot{M}_{a}(T_{vis},~\beta_2)=\dot{M}_{at}(T_{vis1},~\beta_2) + \\
&\ \ \ \frac{T_{vis}-T_{vis1}}{T_{vis2}-T_{vis1}}(\dot{M}_{at}(T_{vis2},~\beta_2)
	- \dot{M}_{at}(T_{vis1},~\beta_2))
\end{split}
\end{equation}.
Then, the second linear interpolation is applied to find the viscous-delayed accretion rates with input 
$T_{vis}$ and with input $\beta$ by
\begin{equation}
\begin{split}
&\dot{M}_{a}(T_{vis},~\beta)=\dot{M}_{a}(T_{vis},~\beta_1) + \\
&\ \ \ \frac{\beta-\beta_1}{\beta_2-\beta_1}(\dot{M}_{a}(T_{vis},~\beta_2)
	- \dot{M}_{a}(T_{vis},~\beta_1))
\end{split}
\end{equation}.
Applications of the linear interpolations can save about one tenth of the running time for the fitting 
procedure to describe the observed long-term light curves, comparing with the running time for the fitting 
procedure considering the integral equation (3) to determine the $\dot{M}_{a}(T_{vis},~\beta)$.

	Third, for a common TDE case with $M_{\rm BH}$ and $M_{*}$ different from $10^6{\rm M_\odot}$ and 
$1{\rm M_\odot}$, the actual viscous-delayed accretion rates $\dot{M}$ and the corresponding time 
information in observer frame are created by the following scaling relations as shown in 
\citet{gm14, mg19},
\begin{equation}
\begin{split}
&\dot{M} = M_{\rm BH,~6}^{-0.5}~\times~M_{\star}^2~\times~
	R_{\star}^{-1.5}~\times~\dot{M}_{a}(T_{vis},~\beta) \\
&t_m = (1+z)\times M_{\rm BH,~6}^{0.5}~\times~M_{\star}^{-1}\times
	R_{\star}^{1.5}~\times~t_{a}(T_{vis},~\beta)
\end{split}
\end{equation},
where $M_{\rm BH,~6}$, $M_{\star}$, $R_{\star}$ and $z$ represent central BH mass in unit of ${\rm 10^6M_\odot}$, 
stellar mass and radius in unit of ${\rm M_\odot}$ and ${\rm R_{\odot}}$, and redshift of host galaxy of a TDE, 
respectively. And the mass-radius relation well discussed in \citet{tp96} is accepted for main-sequence stars.

	Fourth, based on the calculated time-dependent templates of accretion rate $\dot{M}(t)$, the time 
dependent emission spectrum $f_\lambda(t)$ can be calculated through the modeled simple black-body photosphere 
model discussed in \citet{gr13, mg19}
\begin{equation}
\begin{split}
f_\lambda(t)~&=~\frac{2\pi~Gc^2}{\lambda^5}\frac{1}{exp(\frac{hc}{k\lambda~T_p(t)})-1}(\frac{R_P(t)}{Dis})^2 \\
T_p(t)~&=~(\frac{\eta~\dot{M}(t)c^2}{4\pi\sigma_{SB}R_{p}^{2}(t)})^{1/4} \\
R_p(t)~&=~R_0~a_p~(\frac{\eta~\dot{M}(t)c^2}{1.3\times10^{38}M_{BH}})^{l_p} \\
a_p~&=~(GM_{BH}\times\frac{t_p}{\pi})^{1/3}
\end{split}
\end{equation}
with $Dis$ as the distance to the earth calculated by redshift, $k$ as the Boltzmann constant, $T_p(t)$ and 
$R_p(t)$ as the time-dependent effective temperature and radius of the photosphere, respectively,  $\eta$ as 
the energy transfer efficiency smaller than 0.4, $\sigma_{SB}$ as the Stefan-Boltzmann constant, $t_p$ as 
the time information of the peak accretion rate. Then, based on the calculated time dependent $f_\lambda(t)$ 
in observer frame and the wavelength ranges $\lambda_{u,~g,~r,~i,~z}$ of SDSS $ugriz$ filters, the SDSS 
$ugriz$-band time dependent luminosities as discussed in \citet{md15} can be calculated by 
\begin{equation}
L_{u,~g,~r,~i,~z}(t)~=~\int_{\lambda_{u,~g,~r,~i,~z}}f_\lambda(t)d\lambda~\times~4~\pi~\times~Dis^2
\end{equation}
with $Dis$ as the distance to the earth calculated by redshift. Here, not magnitudes but luminosities are 
calculated, mainly because that the collected light curves of SDSS J0159 are the time dependent $ugriz$-band 
luminosities.

	Finally, the theoretical TDE model expected time dependent optical light curves $L_{(u,~g,~r,~i,~z)}(t)$ 
can be described by the following seven model parameters, the central BH mass $\log(M_{\rm BH}/10^6M_\odot)$, 
the stellar mass $\log(M_\star)$ (corresponding stellar radius $R_\star$ calculated by the mass-radius 
relation), the energy transfer efficiency $\log(\eta)$, the impact parameter $\log(\beta)$, the viscous 
timescale $\log(T_{vis})$, the parameters $l_p$ and $R_0$ related to the black-body photosphere model. 
Meanwhile, the redshift 0.312 of SDSS J0159 is accepted. Then, through the through the well-known maximum 
likelihood method combining with the Markov Chain Monte Carlo (MCMC) technique \citep{fd13}, the $ugriz$-band 
light curves of SDSS J0159 can be well described, with accepted prior uniform distributions and starting values 
of the model parameters listed in Table~1. Meanwhile, the available BH masses and stellar masses are the 
ones leading the  determined tidal disruption radius $R_{\rm TDE}$,
\begin{equation}
	\frac{R_{\rm TDE}}{R_{\rm s}} = 5.06\times(M_\star)^{-1/3}(\frac{M_{\rm BH,~6}}
	        {10})^{-2/3}\times R_\star > 1
\end{equation},
to be larger than event horizon of central BH ($R_{\rm s}=2GM_{\rm BH}/c^2$).

\begin{figure*}
\centering\includegraphics[width = 18cm,height=18cm]{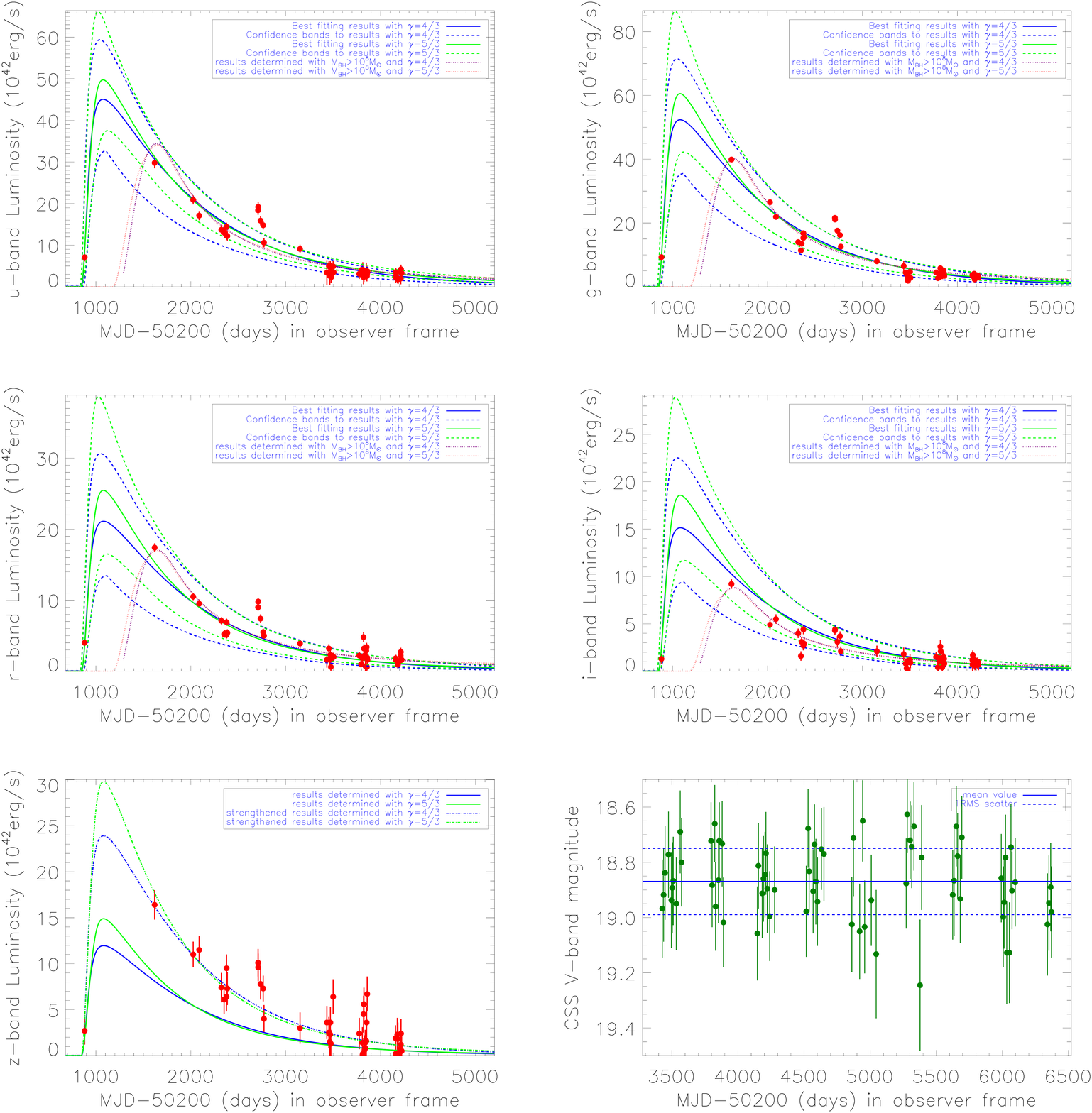}
\caption{The first five panels show theoretical TDE model determined the best fitting results to the SDSS 
$ugriz$-band light curves (solid circles plus error bars in red) of SDSS J0159. In the panels for the $ugri$-band 
light curves, as shown legend in top right corner, solid blue line and dashed blue lines show the best fitting 
results and the corresponding confidence bands calculated by uncertainties of the model parameters determined 
by applications of TDE model with $\gamma=4/3$, solid green line and dashed green lines show the best fitting 
results and the corresponding confidence bands determined by applications of TDE model with $\gamma=5/3$, dotted 
line in purple and dotted line in pink show the TDE model determined descriptions to the light curve by applications 
of prior distribution of central BH mass lager than $10^8{\rm M_\odot}$ in the TDE model with $\gamma=4/3$ and 
$\gamma=5/3$, respectively. In the bottom left panel for the $z$-band light curve, as shown legend in top right 
corner, solid blue line and solid green line show the results determined by applications of TDE model with 
$\gamma=4/3$ and with $\gamma=5/3$, respectively, dot-dashed blue line and dot-dashed green line show the 
corresponding strengthened model results by the factor of 2, which can be applied to well describe the $z$-band 
light curve. Bottom right panel shows the CSS V-band light curve, with horizontal solid and dashed 
lines show the mean value and corresponding 1RMS scatter.}
\label{lmc}
\end{figure*}

\begin{figure*}
\centering\includegraphics[width = 14cm,height=20cm]{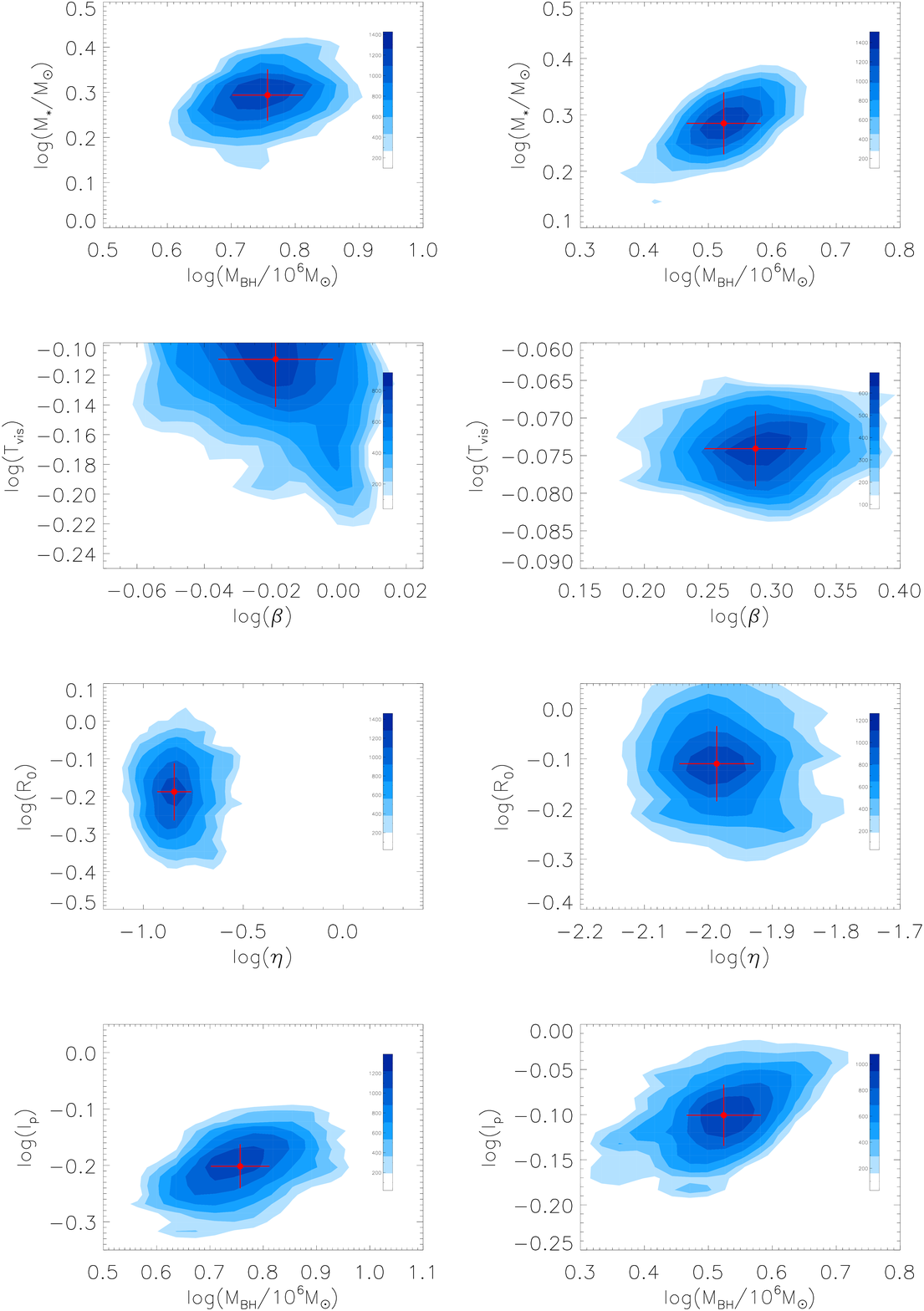}
\caption{MCMC technique determined two dimensional projections of the posterior distributions of the seven 
model parameters. In each left panel, contour represents the results for the model parameters determined by 
TDE model with $\gamma=4/3$. In each right panel, contour represents the results for the model parameters 
determined by TDE model with $\gamma=5/3$. In each panel, solid circle plus error bars in red show the final 
accepted values and the corresponding uncertainties of the model parameters.}
\label{par}
\end{figure*}

\section{Main results and necessary discussions}

	The observed light curves (the time dependent luminosities) in SDSS $ugriz$ bands for SDSS J0159 
are shown in Fig.~\ref{lmc} with the data points are directly collected from Table~2\ in \citet{md15} 
after subtractions of host galaxy contributions. This is the main reason why the time dependent luminosities 
rather than the apparent magnitudes are collected, because the host galaxy contributions have been clearly 
removed from the time dependent luminosities. Then, based on the discussed method and fitting procedure in 
section above, the observed $ugriz$-band light curves of SDSS J0159 are simultaneously described by the 
theoretical TDE model with seven model parameters. Fig.~\ref{lmc} shows the best fitting results and the 
corresponding confidence bands after considerations of the uncertainties of model parameters based on the 
theoretical TDE model. 

	It is clear that besides the $z$-band light curve statistically lower than the theoretical TDE model 
expected values, the other $ugri$-bands light curves can be well described by the theoretical TDE model. 
However, as simply discussed in \citet{gr13, mg19}, the simple black-body photosphere model can be well 
applied to describe optical band variabilities, considering the probable additional contributions of NIR 
emissions from dust clouds, the simple black-body black-body photosphere model is not preferred to describe 
the $z$-band variabilities. Certainly, a simply strengthened TDE model expected theoretical $z$-band light 
curve by a factor 2 (dot-dashed lines in the last panel of Fig.~\ref{lmc}) can be well applied to describe 
the observed $z$-band light curve. Furthermore, as well discussed in \citet{lu16, jn17, jn21}, there should 
be apparent IR contributions to $z$-band light curve of SDSS J0159 at $z=0.312$, considering that radiation 
photons from the central TDE are absorbed by dust grains and then re-radiated in the infrared band, similar 
as discussed in \citet{gr13, mg19}. In other words, if the additional contributions were removed, the $z$-band 
light curve obey the same TDE expected variability properties as those in $ugri$-bands.

	Fig.~\ref{par} shows the MCMC technique determined two dimensional posterior distributions in 
contours of the seven model parameters. The accepted values of the model parameters are listed in Table~1. 
The central BH masses $M_{BH}(10^6{\rm M_\odot})$ in SDSS J0159 are about $\sim5.7_{-0.7}^{+0.8}$ and 
$\sim3.3_{-0.4}^{+0.5}$ by applications of TDE model with $\gamma=4/3$ and with $\gamma=5/3$, respectively. 
Moreover, comparing with the reported parameters for the TDE candidates in \citet{mg19}, the reported 
values of the model parameters are common values among the TDE candidates.

	Before proceeding further, it is necessary to consider whether central BH with mass around 
$10^8{\rm M_\odot}$ (the virial BH mass) can also lead to accepted descriptions to the $ugri$-band 
light curves\footnote{Considering probable contributions from dust emissions, the $z$-band light curve 
is not considered here.} of SDSS J0159. Therefore, besides the best fitting results determined through 
the seven model parameters as free parameters with prior uniform distributions, the TDE model and the 
fitting procedure is running again, but with the central BH mass having the prior uniform distribution 
limited from $1\times10^8{\rm M_\odot}$ to $5\times10^8{\rm M_\odot}$ and the other model parameters 
having the same prior uniform distributions as what have been applied above. Then, the TDE model 
determined descriptions are shown as dotted purple lines and dotted pink lines in the first four 
panels of Fig.~\ref{lmc}, which are worse to explain the rise-to-peak trend. However, if 
the first data point shown in Fig.~\ref{lmc} (MJD-50200=880) was not related to the expected central 
TDE (i.e., starting time of the expected central TDE was later than MJD-50200=880), the determined 
descriptions with accepted virial BH mass could be also well accepted. However, after considering the 
following two points, the data point at MJD-50200=880 is considered to be related to central transient 
event. First, as listed values in Table~2\ in \citet{md15}, the luminosity at MJD-50200=880 is at least 
about 2 times stronger than the data points at late times of the event with MJD-50200 later than 4000, 
indicating that the first data point at MJD-50200=880 related to central event is preferred in SDSS 
J0159. Second, the CSS \citep{dr09} V-band photometric light curve at quiescent state of SDSS J0159 is 
collected and shown in bottom right panel of Fig~\ref{lmc}. The CSS V-band light curve has standard deviation 
about 0.12mag, leading corresponding luminosity variability to be about 11.2\%, quite smaller than the 
luminosity ratio about 2 of the first data point at MJD-50200=880 to the data points at late times.  
Therefore, the model with central BH mass larger than $10^8{\rm M_\odot}$ are not preferred in SDSS J0159.

\begin{figure*}
\centering\includegraphics[width = 16cm,height=10cm]{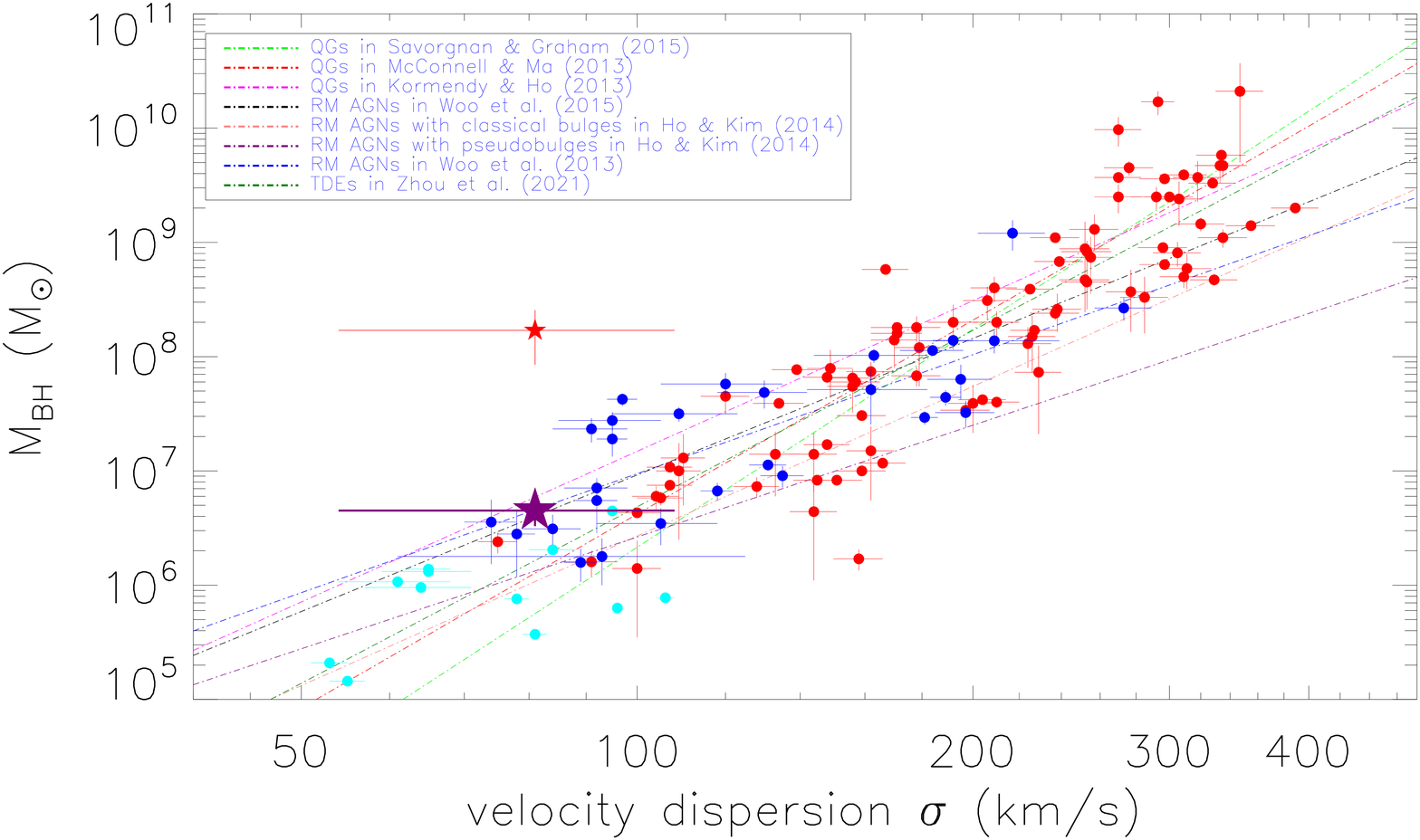}
\caption{On the correlation between stellar velocity dispersions and BH masses. Solid five-point-star 
in purple shows the BH mass of SDSS J0159 determined by theoretical TDE model and the stellar velocity 
dispersion measured in the manuscript, solid five-point-star in red shows the virial BH mass of SDSS 
J0159 as discussed in \citet{lc15, md15, zh19} and the velocity dispersion measured in \citet{zh19}. 
Dot-dashed lines in different colors listed in the legend in top left corner represent the M-sigma 
relations reported in the literature through the quiescent galaxies (QGs) and/or the RM (Reverberation 
Mapped) AGNs with classical/pseudobulges and/or the TDEs. And, solid circles in red, in blue and in cyan 
show the values for the 89 quiescent galaxies from \citet{sg15}, the 29 RM AGNs from \citet{wy15} and 
the 12 TDEs from \citet{zl21}, respectively.}
\label{msig}
\end{figure*}

\begin{figure}
\centering\includegraphics[width = 8cm,height=5.5cm]{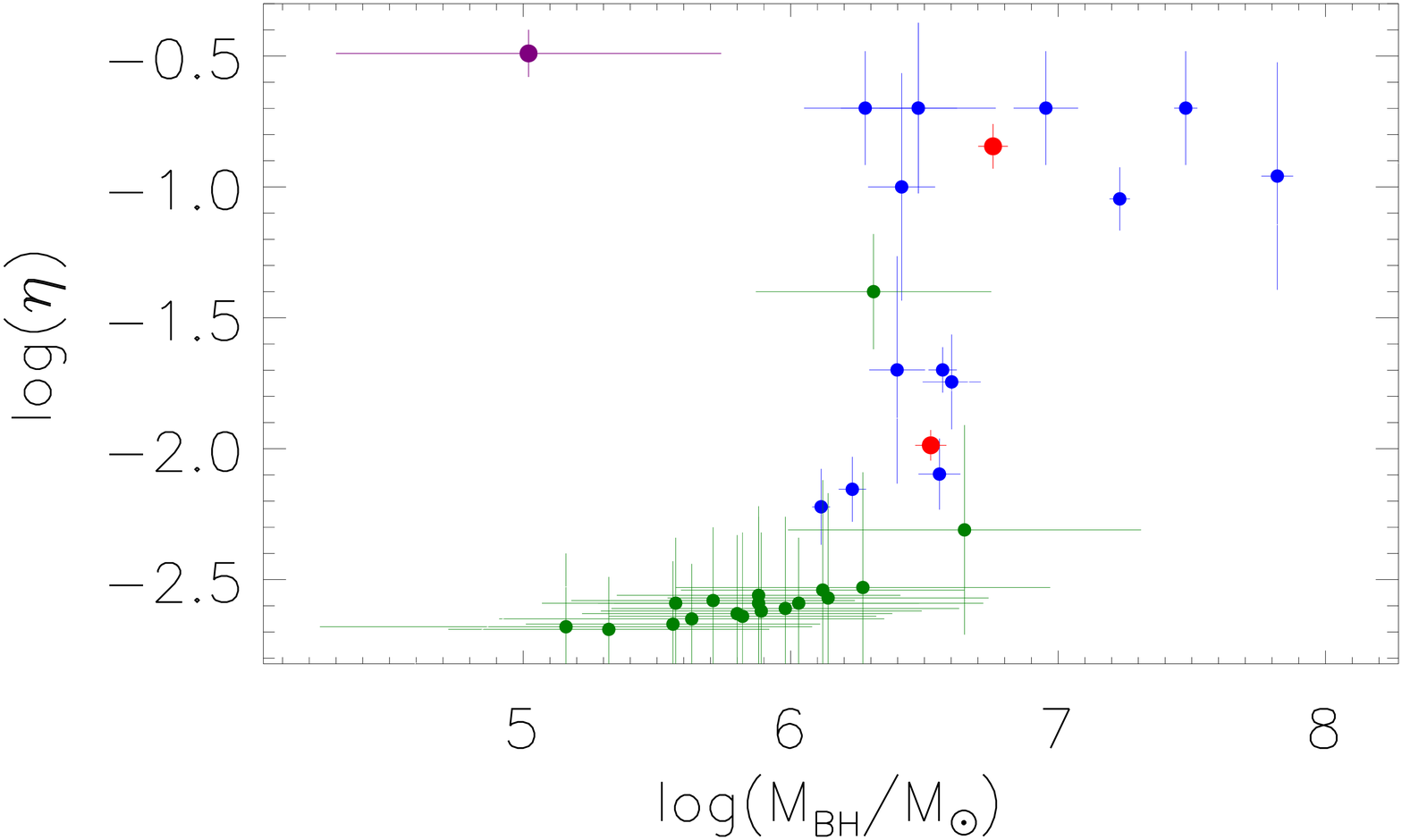}
\caption{On the dependence of energy transfer efficiency $\eta$ on central BH mass of the reported TDE 
candidates in \citet{mg19, zl21}. Symbols in blue represent the results collected from Table~2\ in 
\citet{mg19}, and symbols in dark green represent the results collected from Table~3\ in \citet{zl21}. 
The solid circle plus error bars in purple show the results for the {\it swift} J2058.4+0516 determined 
in \citet{zh22}. The two solid circles in red show the determined results by model with $\gamma=4/3$ 
and $\gamma=5/3$ in the manuscript, respectively.
}
\label{eff}
\end{figure}

\begin{table}
\centering
\caption{Parameters of TDE models for SDSS J0159}
\begin{tabular}{lcrll}
\hline\hline
	parameter & prior & p0 & value$^a$ & value$^b$ \\
\hline
$\log(M_{\rm BH,~6})$ & [-3, 3] & 0. & $0.756\pm0.055$ & $0.524\pm0.058$ \\
$\log(M_\star/M_\odot)$ & [-2, 1.7] & 0. & $0.294\pm0.057$ & $0.285\pm0.055$ \\
$\log(\beta)(4/3)$ & [-0.22, 0.6] & 0. & $-0.019\pm0.017$ & \dots \\
$\log(\beta)(5/3)$ & [-0.3, 0.4] & 0. & \dots & $0.287\pm0.042$ \\
$\log(T_{vis})$ & [-3, 0] & -1. & $-0.109\pm0.032$ & $-0.074\pm0.005$\\ 
$\log(\eta)$  & [-3, -0.4] & -1. & $-0.845\pm0.085$ & $-1.986\pm0.058$ \\
$\log(R_{0})$ &  [-3, 3] & 1. & $-0.187\pm0.076$ & $-0.109\pm0.075$ \\
$\log(l_{p})$  & [-3, 0.6] & 0. & $-0.202\pm0.039$ & $-0.101\pm0.034$ \\
\hline\hline
\end{tabular}
Notes: The first column shows the applied model parameters. The second column shows limitations of the prior 
uniform distribution of each model parameter. The third column with title "p0" lists starting value of each 
parameter. The fourth column with column title marked with $^a$ means the values of model parameters for the 
TDE model with $\gamma=4/3$. The fifth column with column title marked with $^b$ means the values of model 
parameters for the TDE model with $\gamma=5/3$. 
\end{table}

	Based on the results shown in Fig.~\ref{lmc} and posterior distributions of model parameters shown 
in Fig.~\ref{par}, there are two interesting points we can find. First, the observed light curves can be well 
described by the theoretical TDE model, strongly indicating a TDE around central BH of SDSS J0159 as suggested 
in \citet{md15}. Second, different model parameters can lead to observed light curves being well described by 
TDE models, however, the different model parameters in TDE models leading to much different properties on 
features around peak. In other words, once there were observed high quality light curves with apparent features 
around peak intensities, the sole TDE model can be determined. Fortunately, due to the similar BH masses in 
the TDE models with different polytropic indices, it is not necessary to further determine which TDE model, 
the TDE model with polytropic index $\gamma=4/3$ or the model with $\gamma=5/3$, is preferred in SDSS J0159, 
because only BH mass properties are mainly considered in the manuscript.

	Moreover, based on the listed TDE model determined parameters (logarithmic values) in Table~1, the 
stellar mass of the tidally disrupted main-sequence star is about $M_\star\sim1.97{\rm M_\odot}$ 
($M_\star\sim1.93{\rm M_\odot}$) for $\gamma=4/3$ ($\gamma=5/3$), which is out of the transition range between 
$0.3{\rm M_\odot}$ and $1{\rm M_\odot}$ for stars as discussed in \citet{mg19}. Therefore, we do not consider 
hybrid fallback functions that smoothly blend between the 4/3 and 5/3 polytopes, as suggested in \citet{mg19}. 
Meanwhile, as the shown best-fitting results to the $ugri$-bands light curves of SDSS J0159 in Fig.~\ref{lmc}, 
applications of the polytropic index $\gamma=4/3$ and $\gamma=5/3$ individually can lead to well accepted 
descriptions to the observed light curves, re-indicating that it is not necessary to describe long-term 
variabilities of SDSS J0159 with considerations of hybrid fallback functions that smoothly blend between 
the 4/3 and 5/3 polytopes.

	Based on the theoretical TDE model determined BH mass $M_{BH}\sim4.5_{-1.1}^{+1.3}\times10^6{\rm M_\odot}$ 
(the mean value of the two BH masses determined by theoretical TDE models with different polytropic indices), 
we can check the dependence of BH mass on stellar velocity dispersion in SDSS J0159 in Fig.~\ref{msig}. 
Comparing with the reported M-sigma relations for both quiescent galaxies in \citet{mm13, kh13, sg15} and 
in AGN in \citet{hk14, wy15}, the TDE model determined BH mass is well consistent with the M-sigma relation 
expected value in SDSS J0159. Moreover, considering the reported BH masses of TDE candidates in \citet{zl21} 
which we shown as solid cyan circles in Fig.~\ref{msig}, the TDE model determined central BH mass is preferred 
in SDSS J0159, rather than the virial BH mass about two magnitudes higher than the expected value. Furthermore, 
Fig.~\ref{eff} shows the correlation between TDE model determined BH masses $M_{BH}$ and TDE model determined 
energy transfer efficiency $\eta$ of the TDE candidates in \citet{mg19, zl21}. It is clear that the SDSS 
J0159 as a radio quiet object is common as the other TDE candidates in the space of $M_{BH}$ versus $\eta$, 
besides the TDE candidate {\it swift} J2058.4+0516 with relativistic jet related to central TDE \citep{ce12, 
zh22}. Therefore, the BH mass about $10^6{\rm M_\odot}$ in SDSS J0159 is reasonable enough.

	The TDE model determined central BH mass two magnitudes smaller than virial BH mass in SDSS J0159 
provide robust evidence to support that the BLRs in SDSS J0159 include strong contributions of accreting 
debris from central TDE, leading the dynamical properties of disk-like BLRs as discussed in \citet{zh21c}.
And the results in the manuscript to reconfirm that outliers in the space of virial BH masses versus stellar 
velocity dispersions could be better candidates of TDEs.

	Furthermore, simple discussions are given on variability properties of broad H$\alpha$, to support that 
the Virialization assumption applied to estimate virial BH mass is not appropriate to be applied to estimate 
central virial BH mass in SDSS J0159. Under the widely accepted virialization assumption for broad line AGN 
as well discussed in \citet{pe04, gh05, vp06, kb07, rh11, sh11, mt22}:
\begin{equation}
M_{{\rm BH}}~\propto~V^2~\times~R_{{\rm BLR}}~\propto~V^2~\times~L^{0.5}
\end{equation}
where $V$, $R_{{\rm BLR}}$ and $L$ mean broad line width, distance of
BLRs (broad emission line regions) to central BH and broad line luminosity (or continuum luminosity), 
after accepted the well known empirical R-L relation ($R_{{\rm BLR}}~\propto~L^{\sim0.5}$) \citep{kas05, bd13}. 
Then for broad H$\alpha$ in multi-epoch spectra of SDSS J0159, we will have
\begin{equation}
	V_{{\rm 1}}^4~\times~L_{{\rm 1}}~=~V_{{\rm 2}}^4~\times~L_{{\rm 2}}
\end{equation},
where suffix 1 and suffix 2 mean parameters from two different epochs. For SDSS J0159 discussed in
\citet{lc15} and \citet{md15}, the widths (full width at half maximum) of broad H$\alpha$ are about 
(3408$\pm$110)~${\rm km/s}$ and (6167$\pm$280)~${\rm km/s}$ in 2000 and in 2010, respectively, the 
luminosities of broad H$\alpha$ are about ${\rm (329\pm11)~\times~10^{40}~erg/s}$ and 
${\rm (143\pm7)~\times~10^{40}~erg/s}$ in 2000 and in 2010, respectively. Thus, the $V^4~\times~L$ 
in 2000 and 2010 are
\begin{equation}
\begin{split}
&(\frac{V_{{\rm 2000}}}{\rm 1000~km/s})^4~\times~\frac{L_{{\rm 2000}}}{\rm 10^{42}~erg/s}
	~\sim~444_{68}^{77}\\
	&(\frac{V_{{\rm 2010}}}{\rm 1000~km/s})^4~\times~\frac{L_{{\rm 2010}}}{\rm 10^{42}~erg/s}
	~\sim~2068_{435}^{523}
\end{split}
\end{equation}.
There are quite larger $V^4~\times~L$ in 2010 than in 2000. Furthermore, if considering serious obscuration 
of the broad Balmer lines in 2010 (no broad H$\beta$ detected in 2010), the luminosity of broad H$\alpha$ 
in 2010 should be larger, leading to more larger $V^4~\times~L$ in 2010. The results indicating some none-virial 
components are actually included in the broad line variabilities, to support our results in the manuscript 
to some extent.

	Before the end of the section, three additional points are noted. First and foremost, as the shown 
results in Fig.~\ref{lmc}, there is a smaller flare around MJD-50200$\sim$2700 in SDSS J0159, which can be 
simply accepted as a re-brightened peak. Among the reported TDEs candidates, ASASSN-15lh reported in \citet{lf16} 
is the known TDE candidate with re-brightened peak in its long-term light curves. So far, it is not unclear 
on the physical origin of re-brightened peaks in TDEs candidates. As discussed in \citet{lf16}, circularization 
could be efficiently applied to explain re-brightened peak in TDE expected light curves. Meanwhile, as discussed in 
\citet{ml15}, re-brightened peak could be expected when a binary star is tidally disrupted by a central BH. 
And as discussed in \citet{ca18}, re-brightened peak could be expected when a star is tidally disrupted by 
a central binary black hole system with extreme mass ratio. Unfortunately, there are not enough information 
to determine which proposed method is preferred to explain the re-brightened peak around MJD-50200$\sim$2700 
in SDSS J0159. Further efforts in the near future are necessary to discuss properties of the re-brightened 
peak in SDSS J0159. And there are no further discussions on the re-brightened peak in SDSS J0159 in the 
manuscript.

	Besides, as the shown and discussed results of the reported optical TDEs candidates in \citet{mg19, 
gs21, vg21}, time durations about hundreds of days in rest frame can be found at 10\% of the peak of the TDE 
expected light curves, quite smaller than the corresponding time duration of about 1900days at 10\% of the peak 
of the light curves of the SDSS J0159 in the rest frame (with redshift z=0.312 accepted). Actually, the longer 
time duration in SDSS J0159 can be naturally explained by the scaled relation in equation (6) with different 
BH mass and stellar mass. As an example, among the reported optical TDEs candidates shown in \citet{mg19}, 
the PTF09ge has time duration about $T_{P}\sim160days$ at 10\% (2.5magnitudes weaker than the peak magnitude) 
of the peak of its light curves in the rest frame. The PTF09ge is collected, mainly due to more clearer and 
smoother optical light curves with larger variability amplitudes shown in Fig.~1\ in \citet{mg19}. To collect 
one another TDE candidate can lead to similar results. Considering PTF09ge with MOSFIT determined BH mass about 
$M_{BH,P}\sim3.6\times10^6{\rm M_\odot}$ and stellar mass about $M_{\star,~P}\sim0.1{\rm M_\odot}$ (corresponding 
stellar radius about $R_{\star,~P}\sim0.12{\rm R_\odot}$) and $\beta_P\sim1.1$ and $\gamma_P=5/3$ listed in 
\citet{mg19}, the expected time duration $T_{S}$ of the TDE expected light curve in SDSS J0159 can be estimated as 
\begin{equation}
T_{S}~\sim~T_{P}~(\frac{M_{BH,S}}{M_{BH,~P}})^{0.5}~
	(\frac{M_{\star,~S}}{M_{\star,~P}})^{-1}~(\frac{R_{\star,~S}}{R_{\star,~P}})^{1.5}~S_\beta
\end{equation}
with $S_\beta$ as the parameter considering effects of different $\beta$ applied in PTF09ge ($\beta_P\sim1.1$) 
and in SDSS J0159 ($\beta_S\sim1.94$). Then, based on the determined parameters listed in Table~1 for SDSS J0159 
with $\gamma_S=5/3$, the central BH mass and stellar parameters are about $M_{BH,S}\sim3.4\times10^6{\rm M_\odot}$ 
and $M_{\star,~S}\sim1.93{\rm M_\odot}$ (corresponding stellar radius about $R_{\star,~S}\sim1.51{\rm R_\odot}$). 
Meanwhile, based on light curves of the created standard templates of time-dependent viscous-delayed accretion 
rates with $\gamma=5/3$ and $M_{BH}=10^6{\rm M_\odot}$ and $M_\star=1{\rm M_\odot}$, time duration at 10\% of 
the peak of the light curve of the viscous-delayed accretion rate with $\beta=1.94$ and $\log(T_{vis}\sim-0.074)$ 
(the values for the SDSS J0159) is about 1280days, about 4.7 times longer than the corresponding time duration 
about 280days with $\beta=1.1$ and $\log(T_{vis}\sim-1.4)$ (the values for the PTF09ge), leading to 
$S_\beta\sim4.6$. Then, based on the parameters for PTF09ge and for SDSS J0159, we can have 
\begin{equation}
	T_{S}~\sim~160~(\frac{3.4}{3.6})^{0.5}~(\frac{0.1}{1.93})~(\frac{1.51}{0.12})^{1.5}~4.6~\sim~1700{\rm days}
\end{equation}
very similar as the time duration 1900days at 10\% of the peak of the TDE expected flare in the SDSS J0159. 
The results strongly indicate that the longer time durations of the TDE expected flare in SDSS J0159 are reasonable.

	Last but not the least, as discussed in \citet{md15}, there are expected AGN activities through applications 
of narrow emission line flux ratios (see their Fig.~3). Therefore, the central TDE in SDSS J0159 is a central TDE 
in AGN host galaxy, quiet different from the vast majority of the reported optical TDEs candidates in quiescent 
galaxies. Actually, there are so-far only hands of TDEs candidates reported in AGN host galaxies. \citet{bn17} have 
reported a TDE candidate in a narrow line Seyfert 1 galaxy of which light curves can be roughly described by 
theoretical TDE model. \citet{yx18} have shown the TDE expected variability pattern in the low-luminosity AGN NGC 
7213. \citet{ll20} have reported a TDE candidate in AGN SDSS J0227. \citet{fg21} have reported TDEs candidates in 
narrow line Seyfert 1 galaxies. More recently, \citet{zs22} have shown the TDE expected variability patterns in a 
narrow line Seyfert 1 galaxy. Meanwhile, \citet{cp19, cp20} have modeled TDEs variabilities in AGN with a pre-existing 
accretion disc, and shown that about 20days-long plateau could be expected around central BH with masses around 
$10^{6-7}{\rm M_\odot}$. Therefore, considering variability properties of the detected optical TDEs candidates in 
AGN host galaxy reported in the literature, TDE expected variability patterns can also be expected in SDSS J0159. 
Even considering the 20days-long plateau feature by simulating results in \citet{cp20}, the plateau has few effects 
on the long-term variabilities of SDSS J0159, because the time duration about 1900days of the light-curves of SDSS 
J0159 is quite larger than 20days. In one word, the expected central TDE in SDSS J0159 is in AGN host galaxy, but 
TDE described variability patterns can be well expected in SDSS J0159.

\section{Whether the long-term variabilities are related to central AGN activities in SDSS J0159?}

	The results and discussions above are mainly based on the fundamental point that the long-term 
variabilities are related to a central TDE in SDSS J0159. Therefore, it is necessary and interesting to 
check probability of AGN activities applied to well describe the long-term variabilities in SDSS J0159. 
Here, in order to compare variability properties between SDSS J0159 and the other normal quasars in the 
literature \citet{mi10}, rather than the luminosity light curve but the photometric light curve is mainly 
discussed in the section.

\begin{figure*}
\centering\includegraphics[width = 18cm,height=5.5cm]{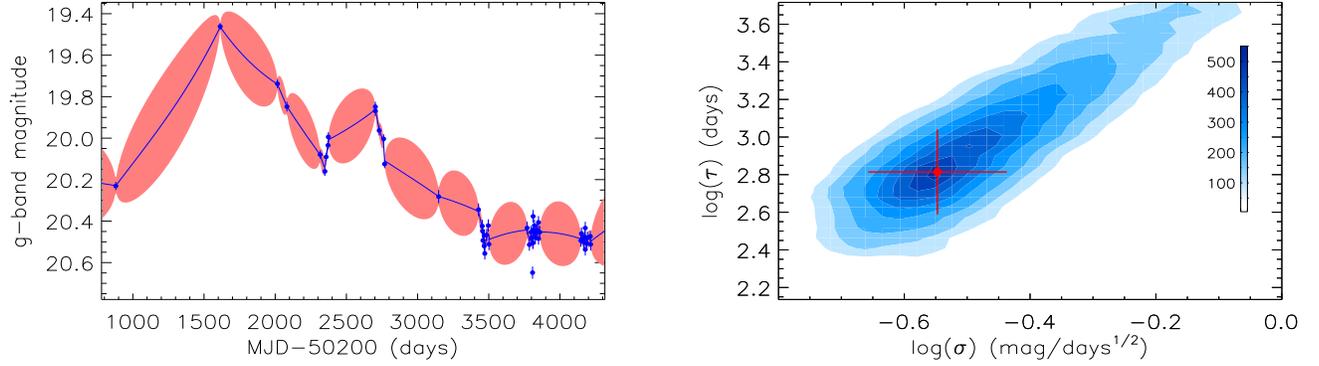}
\caption{Left panel shows the JAVELIN code determined best descriptions to the long-term photometric 
$g$-band (solid circles plus error bars in blue) light curve of SDSS J0159. Solid blue line and area 
filled in pink show the best descriptions and corresponding $1\sigma$ confidence bands to the $g$-band 
luminosity light curve. Right panel shows the MCMC technique determined two-dimensional posterior 
distributions in contour of $\ln(\tau/days)$ and $\ln(\sigma/(mag/days^{0.5}))$, with solid circle plus 
error bars in red showing the accepted values and $1\sigma$ uncertainties of $\ln(\tau)$ and $\ln(\sigma)$. 
}
\label{drw}
\end{figure*}

	The known Continuous AutoRegressive (CAR) process and/or the improved Damped Random Walk (DRW) process 
can be applied to describe intrinsic AGN activities, as well discussed in \citet{kbs09, koz10, mi10, zk13, zk16, 
zh17, mv19, sr22}. Here, based on the DRW process, the collected photometric $g$-band light curve from Stripe82 
database \citep{bv08, ti21} can be described by the public code JAVELIN (Just Another Vehicle for Estimating 
Lags In Nuclei) \citep{koz10, zk13}, with two process parameters of intrinsic characteristic variability amplitude 
and timescale of $\sigma$ and $\tau$. The best descriptions are shown in left panel of Fig.~\ref{drw}. And 
corresponding MCMC technique determined two dimensional posterior distributions of $\sigma$ and $\tau$ are shown 
in right panel of Fig.~\ref{drw}, with the determined $\log(\tau/days)\sim2.92\pm0.21$ and 
$\log(\sigma/(mag/days^{0.5}))\sim-0.55\pm0.10$, leading $SF_\infty/mag\sim\sigma/(mag/days^{0.5})\times\sqrt{\tau/days}$ 
to be about $\log(SF_\infty/mag)\sim0.85$ in SDSS J0159. Actually, to collect the photometric $uri$-band light 
curves can lead to similar results. Meanwhile, as discussed in \citet{mi10} for SDSS normal quasars in Stripe82 
database, normal SDSS quasars have mean values of $\log(\tau/days)\sim2.4$ and $\log(SF_\infty/mag)\sim-0.7$. 
Therefore, in the space $log(\tau)$ versus $\log(SF_\infty)$, SDSS J0159 is an outlier among the quasars, due to 
its $\log(SF_\infty/mag)$ about 1.5magnitudes larger than the normal quasars. In other words, although the SDSS 
J0159 has its light curves with longer time durations, SDSS J0159 has unique variability properties, quite different 
from the normal SDSS quasars.

	Moreover, besides the discussions in the section to show unique variability properties in SDSS J0159, two 
additional points can be found. First, the first data point at MJD-50200=880 is apparently 0.3magnitudes brighter than 
the data points with MJD-50200 later than 4000, which can be applied as one another evidence to support that the first 
data point at MJD-50200=880\ in Fig.~\ref{lmc} is not one data point similar as the other data points at late times, 
but one data point related to the central TDE, under the assumed TDE to explain variabilities of SDSS J0159, similar as 
discussed results in Section 4. Second, although there are unique variability properties in SDSS J0159 quite different 
from normal quasars, contributions of central AGN activities to the light curves cannot be totally ruled out, such as the 
shown weak AGN activities in eBOSS spectrum of SDSS J0159. Therefore, it is necessary to discuss effects of central AGN 
activities on our final results. Besides the fitting procedure discussed in Section 3, five additional model parameters 
are added to described AGN contributions to the light curves, i.e., the equation (8) in Section 3 is re-written to  
\begin{equation}
\begin{split}
L_{u,~g,~r,~i,~z}(t)~=&~\int_{\lambda_{u,~g,~r,~i,~z}}f_\lambda(t)d\lambda~\times~4~\pi~\times~Dis^2\\
	&~+~L0_{u,~g,~r,~i,~z} 
\end{split}
\end{equation}
with $L0_{u,~g,~r,~i,~z}>0$ as the AGN contributions. Then, the theoretical TDE model with 12 model parameters are applied to 
re-describe the $ugriz$-band luminosity light curves shown in Fig.~\ref{lmc} through the Levenberg-Marquardt least-squares 
minimization technique, leading to the parameters $L0_{u,~g,~r,~i,~z}\sim0$ and similar BH masses as discussed in Section 4, 
because the data points at late times have luminosities around zero as shown in Table~2\ in \citet{md15}. If parameters of 
$L0_{u,~g,~r,~i,~z}$ are not constant values but time-dependent, corresponding fitting results to $ugr$-band 
luminosity light curves could lead to stronger luminosities at data points earlier than MJD-50200=880, quite brighter than 
the data points from POSSII shown in Fig.~1\ in \citet{md15}. Therefore, even there are weak AGN activities, they have few 
effects on our final results through applications of theoretical TDE model in the manuscript.

\section{Main Summary and Conclusions}

	Finally, we give our main summary and conclusions as follows. 
\begin{itemize}
\item Host galaxy spectroscopic features are measured by the simple SSP method applied with 350 stellar 
	templates, to confirm the low total stellar mass of SDSS J0159, through the whole spectroscopic 
	absorption features within rest wavelength range from 3650\AA~ to 7700\AA.
\item Theoretical TDE model can be well applied to describe the $ugriz$-band variabilities of SDSS J1059, 
	leading the central BH mass to be about $M_{BH}\sim4.5_{-1.1}^{+1.3}\times10^6{\rm M_\odot}$, two 
	magnitudes smaller than the virial BH mass in SDSS J0159.
\item Through CAR/DRW process applied to describe long-term light curves of SDSS J0159, comparing variability 
	properties between SDSS J0159 and the normal SDSS quasars in Stripe82 database, SDSS J0159 is an outlier 
	in the space of the process parameter of intrinsic variability timescale $\tau$ versus intrinsic 
	variability amplitude $\sigma$, indicating SDSS J0159 has unique variability properties, quite different 
	from the normal quasars.
\item Theoretical TDE model with the model parameter of central BH mass limited to be higher than 
	$10^8{\rm M_\odot}$ cannot lead to reasonable descriptions to the SDSS $ugri$-band variabilities 
	of SDSS J1059, indicating central BH mass higher than $10^8{\rm M_\odot}$ is not preferred in SDSS 
	J0159.
\item The TDE model determine central BH mass is well consistent with the M-sigma relation expected 
	value through the measured stellar velocity dispersion in SDSS J0159.
\item The outliers in the space of virial BH masses versus stellar velocity dispersions could be better 
	candidates of TDEs.
\end{itemize}

\section*{Acknowledgements}
Zhang gratefully acknowledge the anonymous referee for giving us constructive comments 
and suggestions to greatly improve our paper. Zhang gratefully thanks the research funding support from 
GuangXi University and the grant support from NSFC-12173020. This manuscript has made use of the data from 
the SDSS projects. The SDSS-III web site is \url{http://www.sdss3.org/}. SDSS-III is managed by the 
Astrophysical Research Consortium for the Participating Institutions of the SDSS-III Collaboration. The 
paper has made use of the public code of TDEFIT (\url{https://github.com/guillochon/tdefit}) and 
MOSFIT (\url{https://github.com/guillochon/mosfit}), and use of the MPFIT package 
(\url{http://cow.physics.wisc.edu/~craigm/idl/idl.html}) written by Craig B. Markwardt, and use of the 
python emcee package (\url{https://pypi.org/project/emcee/}). The paper has made use of the 
template spectra for a set of SSP models from the MILES \url{https://miles.iac.es}.


\label{lastpage}
\end{document}